\documentclass[twocolumn, times, tighten]{aastex63}

\usepackage{graphicx,multirow,hyperref,url,color,xspace}
\usepackage{tabularx}
\newcommand{\source}{{NGC 4151}\xspace}
\newcommand{\suzaku}{{\textit{Suzaku}}\xspace}
\newcommand{\nustar}{{\textit{NuSTAR}}\xspace}
\newcommand{\integral}{{\textit{INTEGRAL}}\xspace}
\newcommand{\xmm}{{\textit{XMM}}\xspace}
\newcommand{\xray}{{X-ray}\xspace}
\newcommand{\reflkerr}{{\texttt{reflkerr}}\xspace}
\newcommand{\reflkerrlp}{{\texttt{reflkerr\_lp}}\xspace}
\newcommand{\relxilllpCp}{{\texttt{relxilllpCp}}\xspace}
\newcommand{\compps}{{\texttt{compps}}\xspace}
\newcommand{\nthcomp}{{\texttt{nthcomp}}\xspace}
\newcommand{\xillver}{{\texttt{xillver}}\xspace}
\newcommand{\hreflect}{{\texttt{hreflect}}\xspace}
\newcommand{\ztbabs}{{\texttt{ztbabs}}\xspace}
\newcommand{\zxipcf}{{\texttt{zxipcf}}\xspace}
\newcommand{\partcov}{{\texttt{partcov}}\xspace}
\newcommand{\gabs}{{\texttt{gabs}}\xspace}
\newcommand{\warmabs}{{\texttt{warmabs}}\xspace}

\begin{document}

\title{Relativistic reflection in NGC 4151}
\shorttitle{Relativistic reflection in NGC 4151}

\author{Micha{\l} Szanecki}
\affil{Nicolaus Copernicus Astronomical Center, Polish Academy of Sciences, Bartycka 18, PL-00-716 Warszawa, Poland; \href{mailto:mitsza@camk.edu.pl}{mitsza@camk.edu.pl}}
\author{Andrzej Nied{\'z}wiecki}
\affil{Faculty of Physics and Applied Informatics, {\L}{\'o}d{\'z} University, Pomorska 149/153, PL-90-236 {\L}{\'o}d{\'z}, Poland; \href{mailto:andrzej.niedzwiecki@uni.lodz.pl}{andrzej.niedzwiecki@uni.lodz.pl}}
\author{Andrzej A. Zdziarski}
\affil{Nicolaus Copernicus Astronomical Center, Polish Academy of Sciences, Bartycka 18, PL-00-716 Warszawa, Poland; \href{mailto:aaz@camk.edu.pl}{aaz@camk.edu.pl}}

\shortauthors{Szanecki et al.}


\begin{abstract}
We investigate the \xray spectrum of the Seyfert galaxy \source using the simultaneous \suzaku/\nustar observation and flux-resolved \integral spectra supplemented by \suzaku and \xmm observations. Our best spectral solution indicates that the narrow Fe K$\alpha$ line is produced in Compton-thin matter at the distance of several hundred gravitational radii. In such a model, we find a weak but significant relativistic reflection from a disk truncated at about ten gravitational radii when the source is in bright \xray states. We do not find evidence either for or against the presence of relativistic reflection in the dim \xray state. We also rule out models with \xray emission dominated by a source located very close to the black hole horizon, which was proposed in previous works implementing the lamp-post geometry for this source. We point out that accurate computation of the thermal Comptonization spectrum and its distortion by strong gravity is crucial in applications of the lamp-post geometry to the \nustar data. 
\end{abstract}

\section{Introduction}
\label{sec:intro}

\source is a nearby ($z = 0.0033$), archetypal Seyfert 1 galaxy, hosting a supermassive black hole with the mass estimated to $M \simeq (4$--$5) \times 10^7 M_\sun$ \citep{2006ApJ...651..775B, 2014Natur.515..528H, 2014ApJ...791...37O}. It is among the brightest Seyfert galaxies in the hard \xray range and has been extensively studied by all major \xray missions. This observational effort revealed a significant complexity of its \xray spectrum, which includes a blend of several emission and absorption components. The intrinsic \xray emission seems to be produced by thermal Comptonization \citep[but see][]{2017ApJ...850..141B} and it is accompanied by a strong narrow Fe K$\alpha$ emission line as well as a Compton reflection component \citep[e.g.][]{1996MNRAS.283..193Z,2007A&A...463..903D,lubinski10}. The nuclear radiation is strongly absorbed below $\sim$5 keV by an inhomogeneous material with $N_{\rm H} \sim 10^{23}$ cm$^{-2}$, consisting of multiple layers of a neutral and ionized gas along the line of sight \citep[e.g.][]{2002ApJ...573..505Z, 2008ApJ...679.1128K}, some of which undergo rapid changes on a time-scale of days \citep{2007MNRAS.377..607P}. An additional soft \xray emission, which has been spatially resolved on a hundreds of pc scale \citep[e.g.][]{2001ApJ...563..124Y}, dominates the spectrum below $\sim$2 keV and may be produced by scattering of the central continuum into our line of sight \citep[e.g.][]{1994ApJ...423..621W}.

Studies of relativistic reflection, which could indicate the presence of an optically thick disk at a distance of tens gravitational radii ($R_{\rm g}$) or closer, give inconclusive results. No evidence of a relativistically broadened reflection was found e.g.\ by \cite{2003MNRAS.345..423S}, \cite{2019ApJ...884...26Z}, whereas a weak reflection from a truncated disk was estimated by \cite{2001ApJ...549..891W,2002ApJ...573..505Z}. 
\cite{2012MNRAS.422..129Z} reported measurement of energy-dependent lags with an energy profile resembling a relativistically broadened iron line. However, a variety of different profiles were measured later in \source by \cite{2019ApJ...884...26Z}, who then attributed the lags to variations in the absorber rather than to reverberation of the relativistic reflection.
Then, using a long-look simultaneous \suzaku/\nustar observation, \cite{keck15} claimed evidence for reflection from an untruncated disk with a strongly increased emissivity close to the black hole. Such a steep inner emissivity motivates the lamp-post geometry of \citet{1996MNRAS.282L..53M}, with the \xray source located on the symmetry axis of the system, very low above the black hole. A detailed lamp-post model explaining this observation, again including an untruncated disk whose irradiation is strongly dominated by a source located extremely close to the black hole horizon, was later presented by \cite{beuchert17}. 

However, the computational model used for these lamp-post fitting results involved significant inconsistencies in the implementation of general relativity (GR), see discussion in \cite{2016ApJ...821L...1N,2019MNRAS.485.2942N}. In particular, radiation produced so close to the black hole is strongly redshifted, so the rest-frame spectral cutoff, related to the temperature of the \xray source, occurs at an energy much larger than seen by a distant observer. For the most extreme lamp-post models, e.g.\ that fitted to \source, this implies that the rest-frame electron temperature is much larger than 100 keV. The \nthcomp model \citep{1996MNRAS.283..193Z}, applied in the lamp-post model of \citet{beuchert17}, does not properly reproduce the Comptonization spectrum at such large temperatures.  
The related problems where omitted in their model by neglecting the redshift, which is unphysical.
Furthermore, Comptonization spectra deviate from a power-law at relativistic electron temperatures, so using accurate thermal Comptonization models does not only give a more precise rest-frame temperature, but it may actually invalidate the overall fit.

Here we reconsider the \suzaku/\nustar observation to demonstrate how using the proper GR affects the fitting results. We study how the assumed absorption model and the origin of the narrow Fe K$\alpha$ line affect the inferred parameters of relativistic reflection. We also apply the developed spectral models to other archival data sets including hard \xray observations of \source, as data in this energy range are crucial for constraining the Comptonization process.  Specifically, we use the \integral observations in the dim and bright state \citep{lubinski10}, supplemented by \xmm (bright state) and \suzaku (dim state) observations, which correspond to the extrema of hard X-ray flux measured in \source by contemporary detectors.

In Section \ref{sec:dat_red} we describe the data used in our analysis, in Section \ref{sec:spectral} we present our spectral models and the results of their application, and we discuss them in Section \ref{sect:discuss}.

\section{Data reduction}
\label{sec:dat_red}

We use the following three archival data sets providing a broadband coverage of the \xray spectrum, referred to as the spectral sets N, D and B:

\noindent
N. The simultaneous \suzaku and \nustar observation on 2012 November 11--14. Here we use the same data as \citet{keck15} and \citet{beuchert17}, and apply the same data reduction procedure.

\noindent
D. The stacked \integral observations performed between 2005 April 29 and 2007 May 25 in the 'dim' state of \source \citep[as defined by][]{lubinski10}, i.e.\ at the 18-50 keV ISGRI photon flux below $0.004$ s$^{-1}$ cm$^{-2}$, supplemented by the long-look \suzaku observation on 2006 December 18--21, when the source was in that state. Here we use the same data as \cite{lubinski10}, and we apply the same data reduction procedure.

\noindent
B. The \integral observation during 2003 May 23--28 in the 'bright' state of \source \citep[as defined by][]{lubinski10}, at the 18-50 keV ISGRI flux $\simeq 0.01$ s$^{-1}$ cm$^{-2}$, supplemented by a partially simultaneous \xmm observations performed during 2003 May 25--27, obs.\ id 0143500101 (X1), 0143500201 (X2) and 0143500301 (X3). Here, we use the same data as \citet{lubinski10}, and apply the same data reduction procedure. The observations X1 and X2 have been added, and called X12. Observation X3 is treated separately, see Section \ref{sec:abs}.

\begin{figure}
\centerline{\includegraphics[height=11.cm]{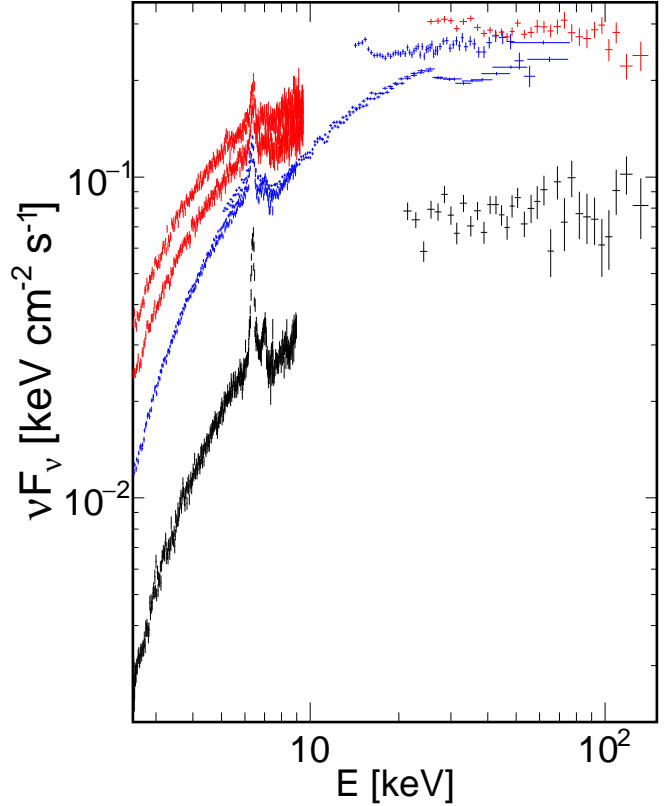}}
\caption{The spectra of \source used in this work. No models have been applied to the data, with the count rates in each energy bin divided by the corresponding effective area. The data points from bottom to top show the spectra D (black), N (blue) and B (red). For B, the X12 and X3 \xmm observations are shown by the lower and upper, respectively, red points in the soft \xray range.
}
\label{fig:counts}
\end{figure}

The \suzaku data have been extracted and reduced using the \texttt{aepipeline} script, with the calibration release from 2015-10-05 for XIS and the calibration release from 2011-09-15 for  HXD. The \nustar data have been reduced with the \texttt{nupipeline} script, using the standard NuSTAR Data Analysis Software NuSTARDAS-v.1.4.1 with the NuSTAR CALDB from 2015-10-08.
The \xmm/EPIC pn data have been reduced using the {\it XMM} Science Analysis Software version 8.0.1 with the standard selection criteria and excluding periods of high or unstable background. The \integral/ISGRI data have been reduced using the Offline Scientific Analysis 7.0 provided by the INTEGRAL Science Data Centre, with the pipeline parameters set to the default values.

\begin{table*}
\caption{\label{tab:summary} Summary of the spectral definitions, cross-calibration constants in our best-fitting models (Section \ref{sec:spectral}), the fit statistics and the model probability determined using the Akaike information criterion (see Section \ref{sect:discuss}). The probabilities are normalized to unity for the sum of models 3, 4 and 6.}
  \resizebox{.95\textwidth}{!}{
\begin{tabularx}{1.05\textwidth}{c|cccc|ccc|ccc}
\hline
       & \multicolumn{4}{c|}{Observations} & \multicolumn{3}{c|}{Cross-calibration constants} & \multicolumn{3}{c}{$\chi^{2}/{\rm DoF}$ \& probability} \\
       & Instrument &  Start date & End date & Exposure [ks] & model 3 & model 4 & model 6 & model 3 & model 4 & model 6 \\
 \hline
 D & \textit{Suzaku} & 2006-12-18 & 2006-12-21 & 125 & 1.0 & 1.0 & 1.0 & 534/407 & 528/410 & 557/417\\[0.1cm]
   & \textit{INTEGRAL} & 2005-04-29 & 2007-05-25 & 651 & $0.56^{+0.11}_{-0.12}$ & $0.70^{+0.19}_{-0.09}$ & $0.84^{+0.17}_{-0.12}$ & $0.002$ & 0.997 & $0.001$ \\[0.1cm]
 \hline
 N & \textit{Suzaku} & 2012-11-11 & 2012-11-13 & 150 & 1.0 & 1.0 & 1.0 & 702/453 & 577/458 & 605/458\\[0.1cm]
   & \textit{NuSTAR} & 2012-11-12 & 2012-11-15 & 140 & $1.03^{+0.03}_{-0.03}$ & $1.03^{+0.03}_{-0.02}$ & $1.03^{+0.03}_{-0.02}$ & $10^{-30}$ & 1.0 & $10^{-6}$\\[0.1cm]
 \hline
 B & \textit{XMM (X12)} & 2003-05-25 & 2003-05-27 & 17 & 1.0 & 1.0 & 1.0 & 697/701 & 685/703 & 690/703\\[0.1cm]
 & \textit{XMM (X3)} & 2003-05-27 & 2003-05-27 & 12 & $1.18^{+0.01}_{-0.02}$ & $1.18^{+0.02}_{-0.02}$ & $1.18^{+0.01}_{-0.02}$ & $0.002$  & 0.918 & $0.08$\\[0.1cm]
 & \textit{INTEGRAL} & 2003-05-23 & 2003-05-28 & 271 & $0.89^{+0.09}_{-0.17}$ & $1.14^{+0.09}_{-0.15}$ & $1.15^{+0.04}_{-0.08}$ & & & \\[0.1cm]
 \hline
 Joint fit & & & & & & & & 1933/1567 & 1790/1573 & 1852/1583\\[0.1cm]
           & & & & & & & & $10^{-34}$ & 1.0  & $10^{-9}$ \\[0.1cm]
 \hline
\end{tabularx}
}
\end{table*}

For the spectral analysis, we use the data in the range of 2.5--9 keV from the XIS front-illuminated (FI, XIS 0 + XIS 3) detectors, 2.5--7.5 from the XIS back-illuminated (BI, XIS 1) detector and 14--60 keV from the PIN, all on board \suzaku, 5--79 keV from the \nustar FPM A and B detectors, 2.5--9.5 keV from the \xmm/EPIC pn and 20--200 keV from the \integral/ISGRI detectors. The data sets from XIS and EPIC pn detectors were binned using the same condition on the minimal signal-to-noise ratio (S/N) of 10. For PIN, both FPM and ISGRI instruments, we use the condition of S/N = 40, 100 and 3, respectively. For XIS, EPIC pn and FPM detectors this gives the width of the spectral bins of about 20 eV around 6.4 keV. 

In the spectral analysis, we apply cross-normalization constants to all detectors relative to the co-added XIS-FI data for spectra D and N and relative to X12 EPIC pn for spectrum B. The cross-normalization constant for \integral/ISGRI and \nustar in our best-fitting spectral models are given in Table \ref{tab:summary}. In all models for spectrum D, the XIS-BI/XIS-FI cross-normalization is fitted at the nominal value, $1.02 \pm 0.01$. In all models for spectrum N, the XIS-BI/XIS-FI cross-normalization is fitted at a lower value, $0.96 \pm 0.01$, which is still reasonable \citep[see][]{keck15}. The cross-normalization factor of PIN relative to XIS-FI is fitted with $1.23 \pm 0.01$ in all cases.

The definition of spectral sets D and B follows from the finding that the intrinsic \xray spectrum of \source changes systematically with the change of its intrinsic \xray luminosity, see \cite{lubinski10}. 
Figure \ref{fig:counts} shows the three spectral sets used in this work. 
The {\it Swift}/BAT photon flux during the \suzaku observation of spectral set D, $\simeq 0.0036 \pm 0.0004$ s$^{-1}$ cm$^{-2}$, was higher than the average ISGRI flux for all D observations, $\simeq 0.0025$ s$^{-1}$ cm$^{-2}$, which justifies the lower than 1 values of the ISGRI normalization relative to \suzaku for this spectrum (see Table \ref{tab:summary}).

\section{Spectral analysis}
\label{sec:spectral}

\subsection{Spectral components}

\source is well known for a significant complexity of its \xray spectrum, which requires several components to be included in the spectral model. We describe here computational models used for these components in our analysis.

\subsubsection{Primary radiation and its reflection from the inner disk}

Our main results are obtained using the relativistic reflection models \reflkerr and \reflkerrlp of \citet{2019MNRAS.485.2942N}. These models involve an accurate computation of the thermal Comptonization spectrum. In particular, they properly describe the distortion of Comptonization spectra in strong gravity. We also recover the lamp-post results of \cite{beuchert17} in the model using \relxilllpCp \citep{2014MNRAS.444L.100D}, where the strong gravity effects on the direct spectrum have been neglected. 

In \reflkerr and \reflkerrlp, the primary radiation is modeled with \compps \citep{ps96}, originally parametrized by the optical depth, $\tau$, electron temperature, $T_{\rm e}$, and the black-body temperature of seed photons, $T_{\rm bb}$. In our implementation of \compps, the photon spectral index, $\Gamma$, can be used instead of $\tau$. All our results correspond to the spherical geometry of the \xray source (\texttt{geometry}=0 in \compps) and we use a constant $kT_{\rm bb}=10$ eV, which is the characteristic disk temperature at the BH mass and the luminosity of \source, see, e.g., \citet{lubinski10}.

In \relxilllpCp, the primary radiation is computed with \nthcomp, parametrized by $T_{\rm e}$ and the photon spectral index, $\Gamma$. This model is inaccurate at high $T_{\rm e}$ and then it is not suitable for computing the spectra affected by strong gravity \cite[see][]{2019MNRAS.485.2942N}. \cite{beuchert17} override this problem by neglecting the effect of the gravitational shift on the \nthcomp spectrum, which then implies that the spectrum in the \xray source frame is strongly blueshifted. However, at the relativistic rest-frame $T_{\rm e}$ implied by this blueshift, the Comptonization spectrum deviates from a power-law, which effect is not reproduced properly by \nthcomp. Comparing the results obtained with \relxilllpCp and \reflkerrlp shows that this is a crucial issue for relativistic reflection models.

The lamp-post models, \reflkerrlp and \relxilllpCp, are parametrized by the black-hole spin parameter, $a$, the \xray source height, $h$, the inner and outer radius of the disk, $r_{\rm in}$ and $r_{\rm out}$ ($h$, $r_{\rm in}$ and $r_{\rm out}$ are in units of $R_{\rm g} = GM/c^2$), and the inclination angle, $i$. The reflection strength (i.e.\ the normalization of the observed reflection with respect to the direct spectral component from the \xray source) is strictly determined for the lamp-post geometry and we do not treat it as a free parameter; however, in the application of \relxilllpCp, we relax this physical constraint in order to reproduce the results of \cite{beuchert17}. In \reflkerr, a phenomenological radial emissivity, $\propto r^{-q}$, is assumed. This version may approximate physical models with various geometries of the \xray source; therefore, here the reflection fraction, $\mathcal{R_{\rm disk}}$, is treated as a free parameter. The definition of $\mathcal{R_{\rm disk}}$ used here follows \citet{2019MNRAS.485.2942N}, i.e.\ $\mathcal{R_{\rm disk}} = 1$ corresponds to the fluxes of radiation incident on the disk and contributing to the directly observed component being equal. \reflkerr is parametrized by $a$, $i$, $q$, $r_{\rm in}$, $r_{\rm out}$ and $\mathcal{R_{\rm disk}}$.

The rest-frame reflection is described by \xillver \citep{garcia13} in \relxilllpCp and by \hreflect \citep{2019MNRAS.485.2942N} in \reflkerr and \reflkerrlp; \hreflect modifies \xillver by using a more accurate model for Compton reflection of \citet{mz95}. In both models, the reflector is characterized by the iron abundance with respect to solar, Z$_{\rm Fe}$, and the ionization parameter, $\xi_{\rm disk}$.

\subsubsection{Absorption}
\label{sec:abs}

A variety of absorption models have been applied to fit the \xray spectra of \source, typically involving at least two absorbers. Here we consider models similar to those used in recent works, including two neutral as well as either one \citep[as in][]{beuchert17} or two \citep[as in][]{keck15, 2019ApJ...884...26Z} ionized absorbers, and allowing them to only partially cover the source. The neutral absorption is described by \ztbabs and modified by \partcov, which converts it to partial covering absorption. The ionized absorption is described using the \zxipcf model. \ztbabs is parametrized by the column density, $N_{\rm H}$, and the redshift which in this component is always set to $z=0.0033$. \partcov is parametrized by the covering fraction, $f_{\rm cov}$. \zxipcf is parametrized by $N_{\rm H}$, $\xi$, $f_{\rm cov}$ and $z$. This model assumes a rather small turbulent velocity of $v_{\rm turb} = 200$ km/s; we have checked that replacing it with \warmabs with $v_{\rm turb} = 3000$ km/s does not affect our results.

Following \citet{beuchert17}, we also considered an absorption line at $E \simeq 8$ keV, modeled with \gabs, which may be due to an ultrafast outflow. We found that it improves the fits to the spectral set N by $\Delta \chi^2 \simeq -15$ for three additional free parameters. However, its inclusion does not affect the fitted parameters of other spectral components. Furthermore, we do not find evidence for such an additional line in either B or D spectrum. Therefore, we neglected it for the results presented below. Instead, taking into account that fast outflows of the ionized absorbers are often observed, we allow the \zxipcf components to be blueshifted in the host galaxy frame by allowing their redshift parameters to vary with $z \le 0.0033$.

In the data set B, the observations X1 and X2 exhibit consistent soft \xray spectra, therefore, we use these observations added together and refer to this stacked spectrum as X12. The observation X3 shows a different absorption level \citep[see][]{lubinski10}. Then, we consider separately X12 and X3 by applying slightly different absorption. In order to minimize the number of free parameters, we vary only $f_{\rm cov}$ of one of the neutral absorbers between these two spectra.

\begin{figure*}
\centerline{\includegraphics[height=8cm]{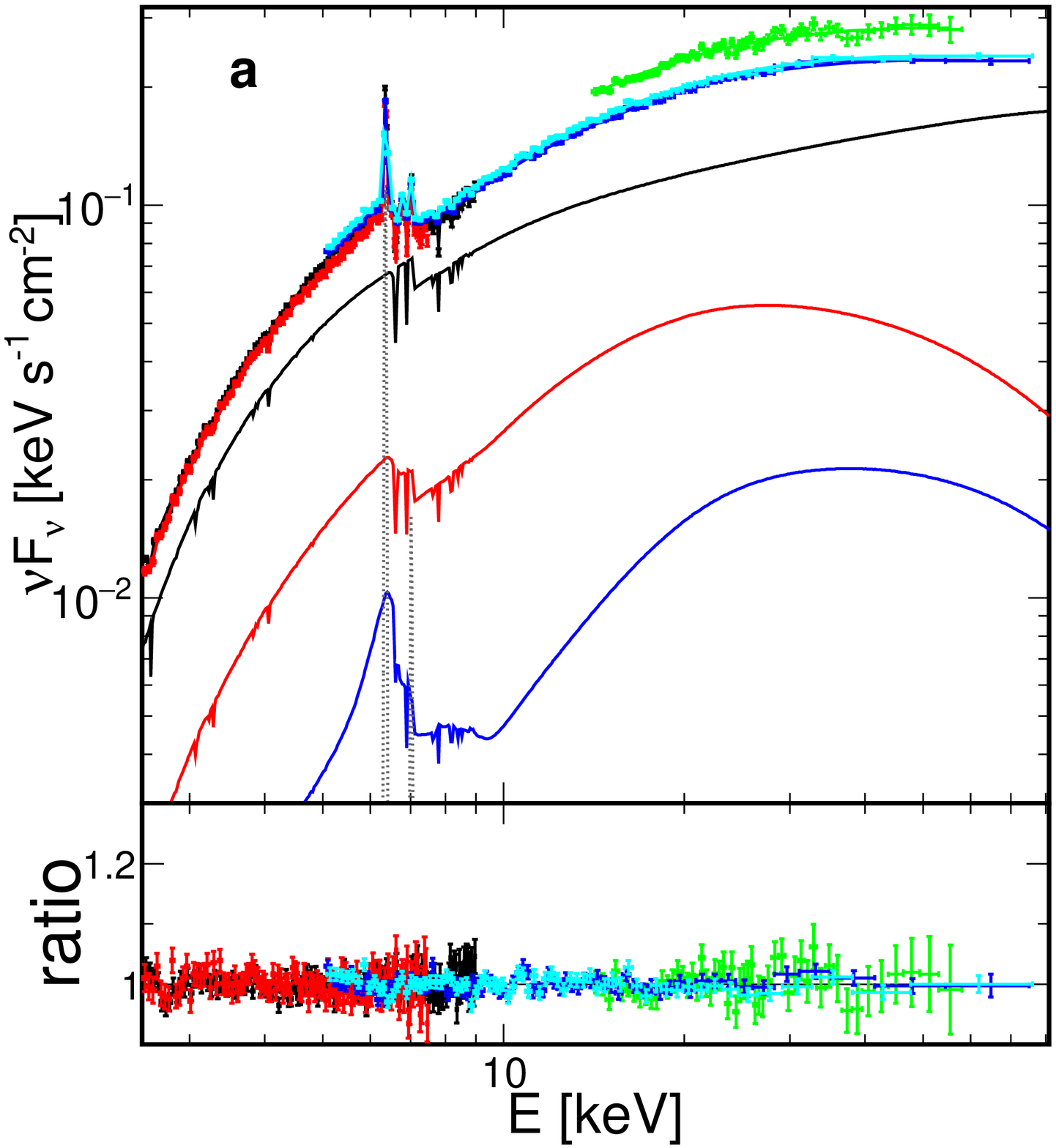}\includegraphics[height=8cm]{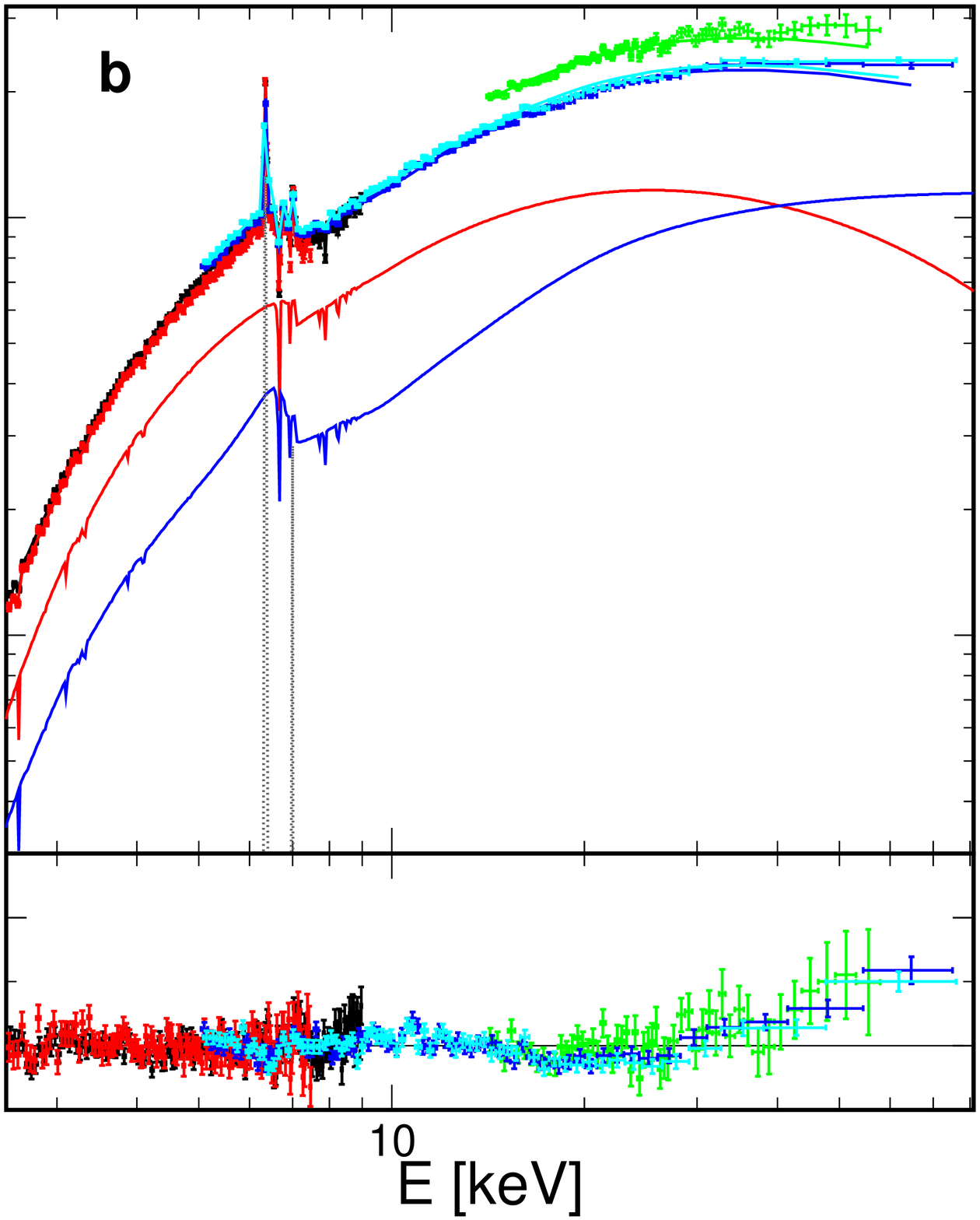}}
\caption{The lamp-post fits to spectrum N. The upper panels show the unfolded data and the model spectral components. The lower panels show the data-to-model ratios. The colors in the data indicate the XIS 0+3 (black), XIS 1 (red), PIN (green), FPM A (blue) and FPM B (cyan). (a) The model 1, following \citet{beuchert17}. The black, red and blue curves show the primary component (\nthcomp) and two reflection components (\relxilllpCp) of the lamps at $h_1=1.17$ and $h_2=15$, respectively. The reflection fractions were free in the fit. (b) The model 2, with the lamppost components modeled with \reflkerrlp. The physical reflection fraction was fixed for both components. The red and blue curves show the total (primary + reflection) spectra of the LP components at $h_1=1.17$ and $h_2=15$, respectively. The gray lines in (a) and (b) show the narrow Fe K$\alpha$ and K$\beta$ lines. All the model components are normalized to the XIS0+3 data. 
}
\label{fig:ngc4151}
\end{figure*}

\subsubsection{Narrow Fe K lines and the distant reflection}
\label{sec:narrow}

\source exhibits a strong, narrow line at $\sim 6.4$ keV, which may be produced, e.g., in (i) a pc-scale ($\sim\! 10^6 R_{\rm g}$ for $M \simeq 5 \times 10^7 M_\odot$) torus \citep{2003MNRAS.345..423S}, (ii) a cold, intermediate disk at a distance of several hundred $R_{\rm g}$ \citep{2018ApJ...865...97M}, or (iii) clouds located $\sim 10^3 R_{\rm g}$ from the center \citep{2019ApJ...884...26Z}. In cases (i) and (iii), the reprocessor can be either Compton-thick or Compton-thin. In cases (ii) and (iii), the line should be (weakly) relativistically distorted. Taking these possibilities into account, we consider the following models: 

(i) {\it Reflection from Compton-thick, static matter.} We assume that the reflector is optically-thick, neutral, distant and static (i.e.\ the reflected spectrum is affected only by the cosmological redshift) and has the same $Z_{\rm Fe}$ as the inner disk. We describe reflection from such a material using \hreflect with $\xi=1$ ${\rm erg\; cm\; s}^{-1}$ and the incident spectrum given by the main Comptonization component of our \reflkerr model. The reflection strength of the distant reflection is a free parameter and has a simple interpretation of $\mathcal{R}_{\rm dist} = \Omega/2\pi$ where $\Omega$ is the solid angle subtended by the reflector at the central \xray source. This model describes a reflector located beyond $\sim 1000 R_{\rm g}$ (i.e.\ the distance at which a relativistic distortion is below the energy resolution of \suzaku or \xmm), e.g.\ a molecular torus, and is represented by model 5 in Section \ref{sec:trunc}. The distant reflector is likely geometrically thick, however, its spatial configuration, in particular the orientation of the reflecting surface with respect to the observer, is uncertain. We assumed $i = 60 \degr$ for this reflection component. We checked that this assumption does not affect our results (the only significant change for using different $i$ in the model 5 concerns the fitted value of $Z_{\rm Fe}$, e.g.\ $Z_{\rm Fe} \simeq 1.8$ for $i = 60 \degr$, whereas $Z_{\rm Fe} \simeq 2.4$  for $i = 30 \degr$).

(ii) {\it Relativistically blurred, Compton-thick reflection.} The reflection here is described by \hreflect, similar as in (i), with an additional weak relativistic blurring described by \texttt{rdblur}, which is the convolution version of the \texttt{diskline} model of \citet{1989MNRAS.238..729F}. \texttt{rdblur} has certain advantages over \reflkerr in modeling the narrow line. First, it allows to compute emission from a disk extending beyond $r_{\rm out} = 10^3$, whereas \reflkerr has a limit of $r_{\rm out} = 10^3$. Second, it has no limit on the disk inclination (which appears to be important for \source), whereas \reflkerr has a limit of $i \ge 9.1 \degr$. For \hreflect we assume $\xi=1$ ${\rm erg\; cm\; s}^{-1}$, we link Z$_{\rm Fe}$ to the inner disk, and we treat the reflection strength, $\mathcal{R}_{\rm dist}$, as a free parameter. This model may describe reflection from the intermediate disk and large values of $\mathcal{R}_{\rm dist}$ may be explained by a warp or a concave structure of the outer disk. Then, we use the same inclination angle in \texttt{rdblur} and \hreflect, but we allow it to be different than the inner disk inclination angle in \reflkerr. We also fix $r_{\rm out}$ of \reflkerr to equal $r_{\rm in}$ of \texttt{rdblur}. The fitted value of this parameter would physically correspond to the radius beyond which the disk is significantly warped. This case is represented by model 6 in Section \ref{sec:trunc}.

(iii) {\it Compton-thin emission.} A Compton-thin material can emit a strong Fe K$\alpha$ line, with the equivalent width (EW) $\sim 100$ eV at the column density $N_{\rm H} \sim 10^{23}$ cm$^{-2}$ \citep[e.g.][]{2003MNRAS.342..422M}, without producing a strong Compton-reflected continuum. We phenomenologically describe the fluorescent iron emission using two Gaussian components representing the Fe K$\alpha$ and K$\beta$ lines with an appropriate branching ratio \citep[with the flux ratio equal to $F_{\beta}/F_{\alpha} = 0.12$, e.g.][]{1978ApJ...223..268B} and the same relative intrinsic widths. We allow these lines to be shifted in energy, assuming that the energy shift is the same for both lines, i.e.\ their centroid energies are linked at $E_{\beta} = (7.06/6.40) E_{\alpha}$. The Gaussians are computed with the \texttt{zgauss} model for $z=0.0033$, so the fitted $E_{\alpha}$ and $E_{\beta}$ give the line energies in the host galaxy frame. This description of the narrow lines is used in all models in Section \ref{sec:lp} and in model 4 in Section \ref{sec:trunc}.

\subsubsection{Soft \xray component}
\label{sec:soft}

Our study is focused on the central \xray source; therefore, we neglect the data below 2.5 keV, at which emission from a large-scale gas starts to dominate. However, we take into account a possible contribution of this extended emission at higher energies and, following \citet{beuchert17}, we assume that it is a nuclear continuum scattered off a distant matter, and model it as an unabsorbed \compps spectrum (or \nthcomp spectrum in the model with \relxilllpCp) with the parameters of the main Comptonization component (except for normalization, which is free).

We find that adding this component is of minor importance. In all cases, its contribution is at most a few per cent at $\la 3$ keV and negligible at higher energies. Furthermore, its inclusion improves our fits by at most $\Delta \chi^2 \simeq -4$ and it does not affect the fitted values of other model parameters.
 
\begin{table*}
\caption{\label{tab:1} The best-fit parameters of the model with two lamppost components above an untruncated disk for spectrum N. The Galactic absorbing column (\texttt{tbabs}), is set to $N_{\rm H} = 2.3 \times 10^{20}\, {\rm cm}^{-2}$. The reflection fractions are kept at the physical values in the models 2 and 3; in the model 1, the two reflection components are fitted freely and the direct radiation is described by an independent \nthcomp component. All models assume $r_{\rm in} = r_{\rm ISCO}$ and $r_{\rm out} = 1000$.
}
\vskip -0.05cm
\newcolumntype{C}[1]{>{\hsize=#1\hsize\centering\arraybackslash}X}%
\newcolumntype{L}{>{\raggedright\arraybackslash}X}%
\resizebox{.95\textwidth}{!}{
\begin{tabularx}{1.17\textwidth}{LC{0.5}C{0.5}C{0.5}C{0.5}C{0.5}}
\hline
\multicolumn{6}{l}{model 1: \texttt{tbabs*(zxipcf*(partcov*ztbabs)*(partcov*ztbabs)*(nthcomp+relxilllpCp+relxilllpCp+zgauss+zgauss)+nthcomp)}} \\
\multicolumn{6}{l}{model 2: \texttt{tbabs*(zxipcf*(partcov*ztbabs)*(partcov*ztbabs)*(reflkerr\_lp+reflkerr\_lp+zgauss+zgauss)+compps)}}\\
\multicolumn{6}{l}{model 3: \texttt{tbabs*(zxipcf*zxipcf*(partcov*ztbabs)*(partcov*ztbabs)*(reflkerr\_lp+reflkerr\_lp+zgauss+zgauss)+compps)}}\\
\hline
 Model component & model 1 & model 2 & \multicolumn{3}{C{1.65}}{model 3}\\
                 & N & N & D & N & B\\
 \hline
 {\bf warm absorber:} &\multicolumn{5}{C{2.75}}{\texttt{zxipcf}}\\
 $N_{\rm H}\;[10^{22}{\rm cm}^{-2}]$ & $4.3^{+2.6}_{-0.4}$ & $1.9^{+0.4}_{-0.2}$ & $64.5^{+50.9}_{-25.6}$ & $3.0^{+0.5}_{-0.4}$ & $64.7^{+34.2}_{-17.6}$ \\[0.1cm]
 $\log_{10}(\xi) $ & $3.37^{+0.03}_{-0.04}$ & $3.53^{+0.03}_{-0.05}$ & $3.2^{+0.2}_{-0.5}$ & $3.8^{+0.1}_{-0.1}$ & $1.0^{+0.3}_{-0.4}$ \\[0.1cm]
 $f_{\rm cov}$ & $0.73^{+0.05}_{-0.04}$ & $1.0^{+0}_{-0.04}$ & $0.17^{+0.06}_{-0.06}$ & $0.31^{+0.69}_{-0.17}$ & $0.19^{+0.28}_{-0.05}$\\[0.1cm]
 $z$ & $0.003^{+0}_{-0.001}$ & $0.003^{+0}_{-0.001}$ & $-0.01^{+0.01}_{-0.01}$ & $-0.009^{+0.007}_{-0.005}$ & $-0.01^{+0.02}_{-0.02}$\\[0.1cm]
 \hline
 {\bf warm absorber:} &\multicolumn{5}{C{2.75}}{\texttt{zxipcf}}\\
 $N_{\rm H}\;[10^{22}{\rm cm}^{-2}]$ & - & - & - & $1.1^{+1.1}_{-0.4}$ & $2.9^{+3.4}_{-1.4}$ \\[0.1cm]
 $\log_{10}(\xi) $ & - & - & - & $4.3^{+0.2}_{-0.1}$ & $4.0^{+0.1}_{-0.2}$ \\[0.1cm]
 $f_{\rm cov}$ & - & - & - & $1.0^{+0}_{-0.16}$ & $1.0^{+0}_{-0.5}$ \\[0.1cm]
 $z$ & - & - & - & $-0.09^{+0.01}_{-0.01}$ & $-0.003^{+0.002}_{-0.003}$ \\[0.1cm]
 \hline
 {\bf neutral absorber:} & \multicolumn{5}{C{2.75}}{\texttt{partcov*ztbabs}} \\
 $N_{\rm H}\;[10^{22}{\rm cm}^{-2}]$ & $12.3^{+1.1}_{-0.1}$ & $8.4^{+0.6}_{-0.1}$ & $15.1^{+1.0}_{-1.1}$ & $11.7^{+0.6}_{-0.4}$ & $8.8^{+0.5}_{-2.1}$ \\[0.1cm]
 $f_{\rm cov}$ & $0.96^{+0.01}_{-0.01}$ & $1.0^{+0}_{-0.27}$ & $1.0^{+0}_{-0.04}$ & $1.0^{+0}_{-0.08}$ & $0.91^{+0.05}_{-0.02}$ \\[0.1cm]
 \hline
 {\bf neutral absorber:} & \multicolumn{5}{C{2.75}}{\texttt{partcov*ztbabs}} \\
 $N_{\rm H}\;[10^{22}{\rm cm}^{-2}]$ & $39.4^{+7.2}_{-1.0}$ & $17.0^{+0.4}_{-0.5}$ & $152.7^{+85.7}_{-18.0}$ & $32.0^{+1.3}_{-1.1}$ & $13.1^{+4.4}_{-5.5}$ \\[0.1cm]
 $f_{\rm cov}$ & $0.28^{+0.05}_{-0.01}$ & $0.49^{+0.21}_{-0.21}$ & $0.34^{+0.08}_{-0.03}$ & $0.33^{+0.01}_{-0.03}$ & $0.26^{+0.18}_{-0.04}$ X12\\[0.1cm]
                                                                                         &  &  &  & &$0.05^{+0.18}_{-0}$ X3\\[0.1cm]
 \hline
 {\bf narrow K$\alpha$:} &\multicolumn{5}{C{2.75}}{\texttt{zgauss}}\\
 $E_\alpha$ [keV] & $6.38^{+0.01}_{-0.01}$ & $6.37^{+0.01}_{-0.01}$ & $6.39^{+0.01}_{-0.01}$ & $6.37^{+0.01}_{-0.01}$ & $6.40^{+0.01}_{-0.01}$ \\[0.1cm]
 $\sigma$ [eV] & $15^{+10}_{-6}$ & $10^{+8}_{-9}$ & $<15$ & $27^{+9}_{-7}$ & $64^{+12}_{-7}$\\[0.1cm]
 $N_\alpha \; [10^{-4}$ph/(cm$^2$s)$]$ & $2.4^{+0.1}_{-0.1}$ & $2.1^{+0.1}_{-0.1}$ & $2.5^{+0.2}_{-0.1}$ & $2.5^{+0.1}_{-0.1}$ & $2.7^{+0.3}_{-0.2}$ \\[0.1cm]
 \hline
 {\bf LP components:} & \texttt{relxilllpCp}& \multicolumn{4}{C{2.2}}{\texttt{reflkerr\_lp}}\\
 $a$ & 0.998 (f) & 0.998 (f) & \multicolumn{3}{C{1.65}}{$0.998^{+0}_{-0.001}$(l)}\\[0.1cm]
 $i\;[\degr]$ & $3.4^{+0.1}_{-0.1}$ & $22.4^{+0.7}_{-0.3}$ & \multicolumn{3}{C{1.65}}{$9.1^{+1.3}_{-0}$(l)}\\[0.1cm]
 $Z_{\rm Fe}$ & $2.2^{+0.1}_{-0.2}$ & $2.0^{+0.1}_{-0.2}$ & \multicolumn{3}{C{1.65}}{$2.6^{+0.1}_{-0.2}$(l)} \\[0.1cm]
 $\log_{10}(\xi_{\rm disk})$ & $3.00^{+0.01}_{-0.01}$ & $3.00^{+0.01}_{-0.01}$ & $3.08^{+0.07}_{-0.04}$ & $3.19^{+0.03}_{-0.01}$ & $3.7^{+0.3}_{-0.2}$ \\[0.1cm]
 $\Gamma$ & $1.72$(f) & $1.70^{+0.01}_{-0.02}$ & $1.42^{+0.06}_{-0.01}$ & $1.64^{+0.04}_{-0.01}$ & $1.47^{+0.05}_{-0.04}$ \\[0.1cm]
 $kT_{\rm e}\;[\rm keV]$& 400(f) & 400(f) & $230^{+63}_{-126}$ & $403^{+4}_{-20}$ & $80^{+30}_{-14}$\\[0.1cm]
 {\it{\bf LP$_{1}$}} & & & &\\
 $N \; [\times 10^{-2}]$ & $3.9^{+0.5}_{-0.1}$ & $1.74^{+0.03}_{-0.05}$ & $0.3^{+0.3}_{-0.1}$ & $1.8^{+0.2}_{-0.1}$ & $0.7^{+0.5}_{-0.5}$ \\[0.1cm]
 $h\, [r_{g}]$ & 1.17 (f) & 1.17 (f) &  $1.17^{+0.12}_{-0.02}$ & $1.33^{+0.02}_{-0.02}$ & $1.6^{+0.2}_{-0.2}$\\[0.1cm]
 {\it{\bf LP$_{2}$}} & & & &\\
 $N \; [\times 10^{-2}]$ & $0.71^{+0.18}_{-0.04}$ & $1.36^{+0.03}_{-0.04}$ & $0.4^{+0.1}_{-0.1}$ & $0.56^{+0.02}_{-0.19}$ & $1.2^{+0.5}_{-0.2}$ \\[0.1cm]
 $h\,[r_{g}]$ & {15 (f)} & {15 (f)} & $61^{+44}_{-14}$ & $18.2^{+4.5}_{-4.2}$ & $5.9^{+1.9}_{-2.5}$\\[0.1cm]
 {\bf primary} & \texttt{nthcomp} & & & & \\
 $N \; [\times 10^{-2}]$ & $5.2^{+0.04}_{-0.03}$ & \multicolumn{4}{C{2.2}}{-}\\
 \hline
 \it{\bf soft component} & \texttt{nthcomp} & \multicolumn{4}{C{2.2}}{\texttt{compps}} \\
 $N \; [\times 10^{-3}]$ & $1.4^{+0.1}_{-0.1}$ & $1.9^{+0.1}_{-0.1}$ & $<1.5$ & $2.9^{+0.1}_{-0.3}$ & $2.1^{+0.9}_{-1.1}$ \\[0.1cm]
 \hline
 $\chi^{2}/{\rm DoF}$ & 607/461 & 928/461 & 534/407 & 702/453 & 697/701\\[0.1cm]
 $\chi^{2}/{\rm DoF}$ & - & - & \multicolumn{3}{C{1.63}}{1933/1567}\\[0.1cm]
 \hline
\end{tabularx}
}\\
\vskip 0.01cm
{\it Notes:} The second, unabsorbed (except for the Galactic ISM) Comptonization (soft) component, \nthcomp in 1 and \compps in 2 and 3, has the spectral index linked to $\Gamma$ of the main Comptonization component (see Section \ref{sec:soft}). The second Gaussian component represents the Fe K$\beta$ line (see Section \ref{sec:narrow}). The normalization, $N$, of Comptonization components give the 1-keV flux in keV cm$^{-2}$ s$^{-1}$. The LP components in 1 neglect the direct component \citep[following][]{beuchert17} but we give this model normalization corresponding to \texttt{fixReflFrac}=1. Parameters denoted with '(f)' were fixed in the fitting.
$\xi$ is given in the unit of ${\rm erg\; cm\; s}^{-1}$. In model 3 parameters denoted with '(l)' are linked across the three spectra. DoF for the individual spectra include these parameters, therefore, their sum is larger than the DoF for the joint fit.
\end{table*}

\subsection{Lamppost with an untruncated disk}
\label{sec:lp}

We present here our lamppost model results. In all lamppost models discussed here, we use the same additional spectral components as \citet{beuchert17}, i.e., a warm absorber an two neutral, partially-covering absorbers (Section \ref{sec:abs}), narrow Fe K$\alpha$ and K$\beta$ lines (Section \ref{sec:narrow}) and an additional Comptonization component attenuated only by the Galactic absorption (Section \ref{sec:soft}).

We first use the lamppost model previously applied to spectrum N (\relxilllpCp) and compare it with an improved model using \reflkerrlp, which allows us to illustrate the importance of the proper modeling of gravitational redshift. Spectrum N was previously analyzed with lamppost models by \citet{keck15} and \citet{beuchert17}. They claimed the presence of an untruncated accretion disk around a rapidly rotating BH, illuminated by two sources, one of them located very close to the BH horizon. These two point-like sources are supposed to approximate a single vertically extended corona, therefore, the parameters of their intrinsic Comptonization spectra are linked. We first assume the parameters of the model fitted in \citet{beuchert17}, including two lamppost (LP) components at $h_1=1.17$ and $h_2=15$, the disk truncated at $r_{\rm in}=r_{\rm ISCO}$, and $a=0.998$. Our aim is to illustrate the GR effects by comparing two models with the same lamppost parameters, therefore, we consider two models with parameters of the LP components fixed at these values. 

In the model 1, we follow \citet{beuchert17}, i.e., we include two lamppost reflection components, LP$_1$ and LP$_2$, described by \relxilllpCp (neglecting their direct components) and one thermal Comptonization component described by an independent \nthcomp spectrum, and we allow for free normalizations of these three spectral components. This model neglects the GR redshift of the primary emission as well as it does not use the physical reflection fractions. In the model 2, we use two \reflkerrlp components including both the primary and reflected radiation and in both we fix the physical reflection fraction. In both models, we also fix $kT_{\rm e}=400$ keV (following \citealt{beuchert17}). 

Our fitting results are shown in Table \ref{tab:1} and Fig.\ \ref{fig:ngc4151}. The model 1 gives a good fit, similar to that found in \citet{beuchert17}. The model 2 gives a significantly worse fit, with $\Delta \chi^2 \simeq 300$, and a strong residual pattern seen in Fig.\ \ref{fig:ngc4151}(b). The discrepancy is related with the redshift of the primary component from $h_1$, taken into account in \reflkerrlp and neglected in the model 1. Allowing $kT_{\rm e}$ to vary in the model 2 does not improve the fit and the fitted value remains $\simeq 400$ keV. This behavior of the model results from properties of thermal Comptonization spectra, which deviate from a power-law shape at large (rest-frame) $T_{\rm e}$. Such deviations appear already at the fitted $kT_{\rm e} \simeq 400$ keV and $\Gamma \simeq 1.7$ (see, e.g., figure 4 in \citealt{2019MNRAS.485.2942N}) and are more pronounced for larger $T_{\rm e}$. Therefore, an increase of $T_{\rm e}$, which could compensate for the redshift, would also lead to departures from a power-law shape and a disagreement with the data. 

We note that the physical reflection fractions used in the model 2 are crucial for the above discrepancy. If we use this model with free reflection fractions (which is unphysical), we find we can obtain a good fit, with $\mathcal{R_{\rm disk}} \simeq 0.8$ for LP$_1$   and  $\mathcal{R_{\rm disk}} \simeq 0.1$ for LP$_2$. It includes a weakly redshifted primary spectrum of LP$_2$ stronger than that of LP$_1$, and completely dominating at high energies. 

Also the physical reflection fraction gives a slightly larger $i$ in the model 2 relative to model 1. The tendency of the increase of $i$ is due to the related decrease of the contribution of the direct component from LP$_1$, which is too strongly redshifted to fit the data. The fitted $i \simeq 22 \degr$ is the largest inclination for which the related change of the line profile can be compensated by the change in absorption.

We now modify model 2 by allowing $a$, $T_{\rm e}$ and both $h$ to vary. We refer to this version as model 3 and we apply it to all three spectral sets, linking $a$, $i$ and Z$_{\rm Fe}$ across them. We also constrained $h \le 2$ for LP$_1$, and we added a second ionized absorber, i.e.\ we now use the same absorption model as for the models with the truncated disk in Section \ref{sec:trunc}. For spectrum D the second warm absorber does not give a statistically significant improvement ($\Delta \chi^2 = -3$ for 4 degrees of freedom, which decreases the model likelihood), therefore, we neglected it. The model 3 results are shown in Table \ref{tab:1}. We see that the parameters fitted for spectrum N are not significantly different from those in the model 2 and the quality of the spectral description is still very poor and much worse than that with the truncated disk models in Section \ref{sec:trunc}. On the other hand, the quality of the model 3 fits to spectra D and B is not so strongly worse than that for the fits with the truncated disk models. In these model 3 fits to spectra D and B, however, the directly observed radiation is dominated by the LP$_2$ component (i.e.\ the one at a larger $h$), in contrast to the model for spectrum N.  Therefore, the redshift of the direct radiation is not significant.

 Then, we modify model 3 by removing the LP$_2$ component and for LP$_1$ we keep the constraint of $h \le 2$, with other assumptions of the model not changed. Even in such a model we do not find a strong systematic disagreement with the data such as that estimated above for spectrum N. Although the fitted temperatures are now very large, $kT_{\rm e} \simeq 600$ keV for D and $kT_{\rm e} \simeq 500$ keV for B, we find that the related distortions of the Comptonization spectrum can be compensated by the change of the cross normalization of the ISGRI data within its uncertainty limit. We conclude that the spectral sets including only the \integral data in the hard \xray range do not allow to rule out spectral solutions with a strongly redshifted thermal Comptonization component in a manner similar to that inferred with the \nustar data. 

\begin{figure}
\centerline{\includegraphics[width=8cm]{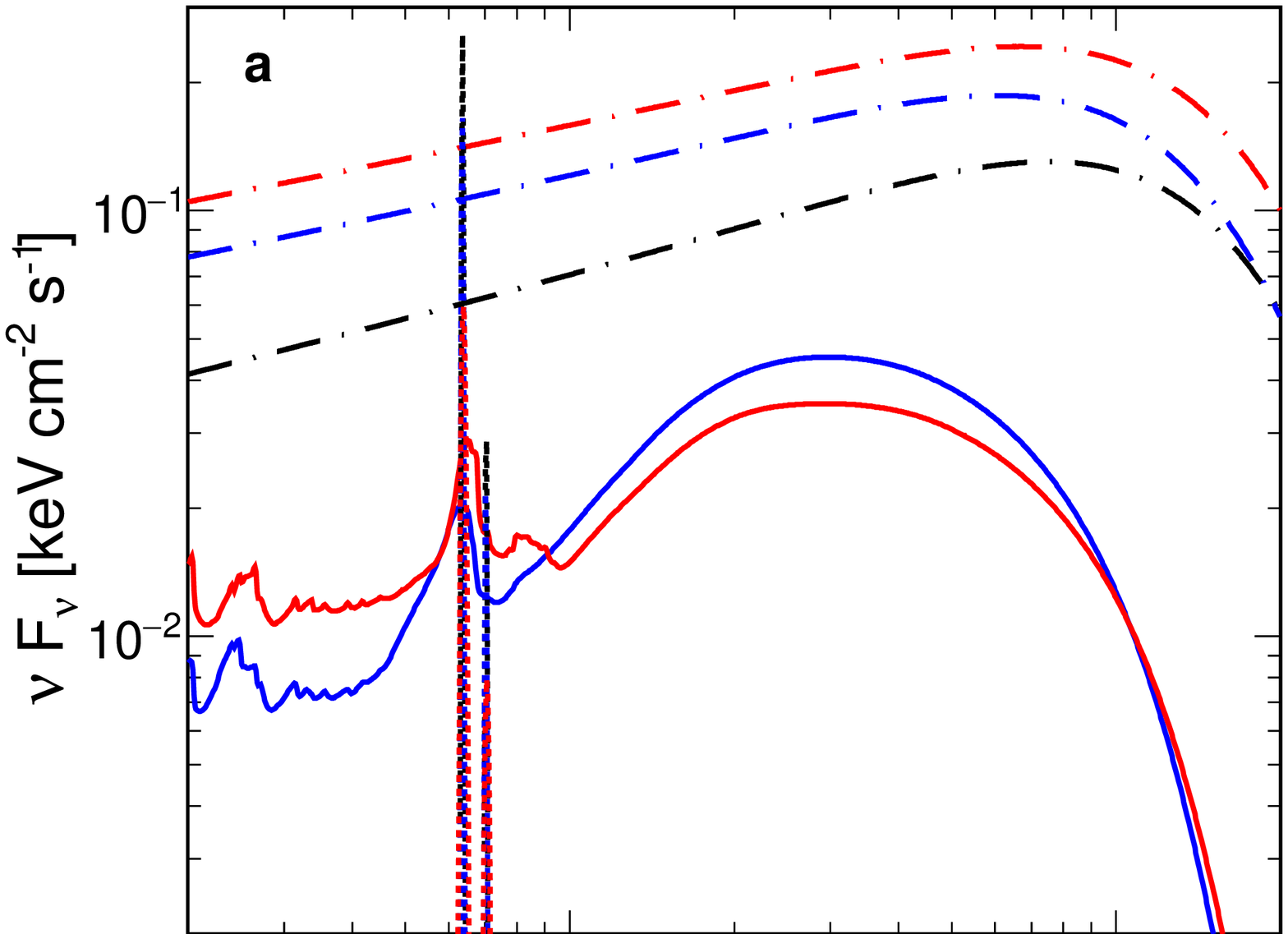}}
\centerline{\includegraphics[width=8cm]{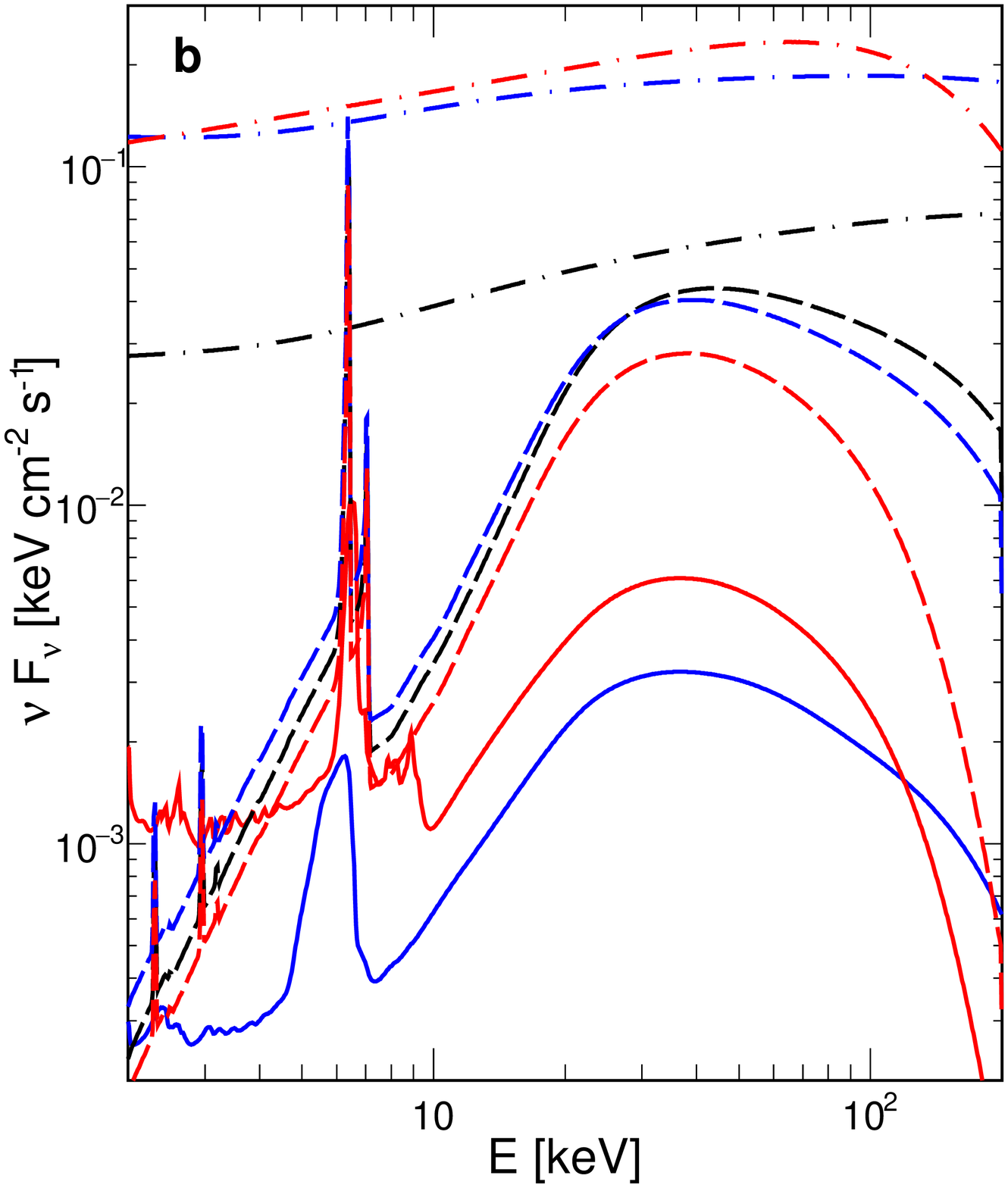}}
\caption{Unabsorbed components of model 4 (a) and 6 (b), fitted to spectrum B (red), N (blue) and D (black). In both panels, the dot-dashed curves show the \compps spectra and the solid curves show the \reflkerr spectra. The dotted lines in (a) show the \texttt{zgauss} components, the dashed curves in (b) show the \texttt{rdblur*hreflect} components. 
}
\label{fig:q3}
\end{figure}

\subsection{Truncated disk}
\label{sec:trunc}

We now analyze the spectra N, D and B using models with a free $r_{\rm in}$. Allowing $r_{\rm in}$ to vary in the lamp-post model, we typically find $r_{\rm in} \ga 7$ and $h < r_{\rm in}$. For such parameters, the disk irradiation is $\propto r^{-3}$. Therefore, for our further analysis we use the \reflkerr model with $q=3$. This model can be considered as an approximation of an inner hot flow irradiating a truncated disk, which we regard as a likely geometry for the fitted parameters. The \reflkerr model gives very similar results to \reflkerrlp except for the rest-frame $T_{\rm e}$ fitted to N, which is higher by a factor $\la 2$ in the lamp-post fits. Also for the fitted values of $r_{\rm in}$, the model spectra are negligibly affected by the spin value. For simplicity we assume $a=0.998$ for the results presented below. 

The unknown origin of the narrow Fe K$\alpha$ line results in the major uncertainty for spectral modeling of \source. We have considered various models as described in Section \ref{sec:narrow}. In model 4 we describe the narrow Fe K lines by the \texttt{zgauss} components and in model 5 we assume that the lines are produced by Compton-thick reflection computed with \hreflect. Model 6 is similar to model 5, but the \hreflect reflection is weakly smeared by \texttt{rdblur}. Each model was fitted jointly to spectrum N, B and D, with $i$ and Z$_{\rm Fe}$ linked across the three spectral sets. For all results presented in this section, we use two neutral and two ionized absorbers, except for the models 5 and 6 for spectrum D, where the second warm absorber as well as the neutral absorbers did not improve the fit.

\begin{figure}
\centerline{\includegraphics[width=9.cm]{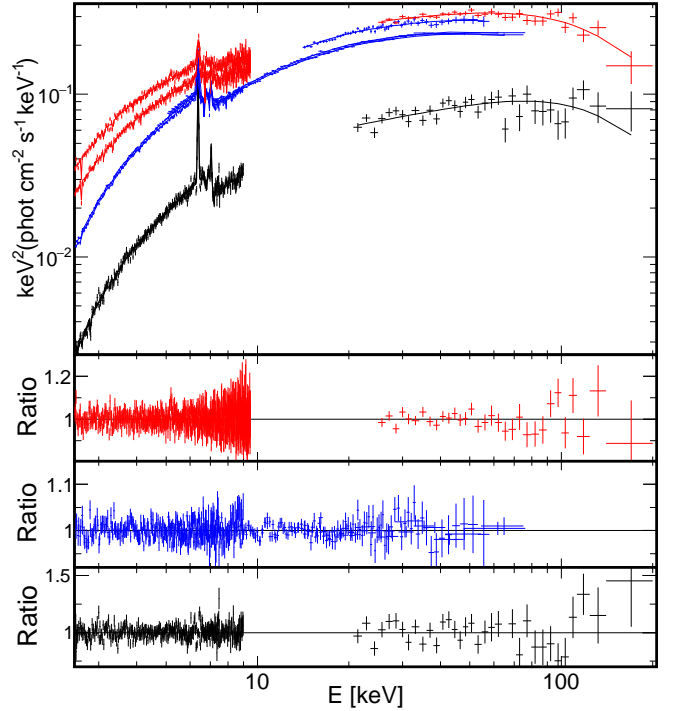}}
\caption{Fits of the model 4 to the spectra D (black), N (blue) and B (red). The upper panel shows the unfolded data and the total model spectra. The lower panel shows the data-to-model ratios For B, N and D from top to bottom.
}
\label{fig:eeuf}
\end{figure}

The results for models 4 and 6 are given in Table \ref{tab:model456} and Figure \ref{fig:q3} shows the main emission components of these fitted models. Figure \ref{fig:eeuf} shows spectra for model 4. For model 5 we found very similar parameters of \reflkerr to those fitted in model 6 (except for a smaller $Z_{\rm Fe} \simeq 1.8$ in model 5) but with a much worse $\chi^{2}/{\rm DoF} = 1921/1587$. The deficiency of model 5 is due to its inability to reproduce the redshifted component of the narrow Fe K$\alpha$ line, which is present in the observed spectra D and N, as indicated e.g.\ by the fitted centroid energies of \texttt{zgauss} in model 4. In spectrum B the narrow line is not redshifted. Remarkably, for spectrum N the same value of the redshift is indicated independently by the \suzaku and the \nustar observations. Fitting these observations separately with model 4, in both cases we found the same $E_{\alpha}=6.38^{+0.01}_{-0.01}$ keV (in the host galaxy frame).

This fitted $E_{\alpha}$ may result from the gravitational redshift if the emission arises at $r \sim 350$. Similar distance of the emission region is indicated by the fitted $r_{\rm in}$ of \texttt{rdblur} in model 6, where also the fitted \texttt{rdblur} inclination angle is $i \simeq 3 \degr$. Also similar results are obtained when the \texttt{zgauss} components in model 4 are replaced by \texttt{diskline}. The fitted \texttt{diskline} parameters are $r_{\rm in} = 400^{+160}_{-100}$ (D) and $225^{+45}_{-35}$ (N), and $i < 8 \degr$.

Model 4 gives a better spectral fit than model 6, which indicates that the narrow Fe K line is produced in a Compton-thin rather than Compton-thick material. Distinguishing between these two cases is crucial for the assessment of the presence of the inner disk. In model 4 we find a significant evidence for reflection, with $\mathcal{R}_{\rm disk} \sim 0.2$, from a disk truncated at $r_{\rm in} \simeq 8$ in N and at $r_{\rm in} \simeq 60$ (but poorly constrained) in B. We checked the significance of inner disk reflection individually for each spectral set. Neglecting the relativistic reflection (i.e.\ setting $\mathcal{R}_{\rm disk}=0$) gives $\Delta \chi^2 = 80$ for N and $\Delta \chi^2 = 16$ for B. For spectrum D we found no significant indication for the inner disk reflection ($\Delta \chi^2 = -4$ for 3 degrees of freedom), therefore, we neglected it.

\begin{table*}
 \caption{The best-fit parameters for models including reflection from a truncated disk, described by \reflkerr with $q=3$ and $a=0.998$. Model 4 assumes $r_{\rm out}=1000$. In model 6, $r_{\rm out}$ of the \reflkerr model equals $r_{\rm in}$ of \texttt{rdblur}. The bottom row gives the contribution of each spectral set to the total $\chi^{2}$ of the joint fit to the three spectra. 
 }
   \newcolumntype{C}[1]{>{\hsize=#1\hsize\centering\arraybackslash}X}%
   \newcolumntype{L}{>{\raggedright\arraybackslash}X}%
   \resizebox{.95\textwidth}{!}{   
   \begin{tabularx}{1.12\textwidth}{LC{0.4}C{0.4}C{0.4}C{0.4}C{0.4}C{0.4}}
 \hline
 \multicolumn{7}{l}{model 4: \texttt{tbabs*(zxipcf*zxipcf*(partcov*ztbabs)*(partcov*ztbabs)*(reflkerr+zgauss+zgauss)+compps)}} \\
 \multicolumn{7}{l}{model 6: \texttt{tbabs*(zxipcf*zxipcf*(partcov*ztbabs)*(partcov*ztbabs)*(reflkerr+rdblur*hreflect)+compps)}} \\
 \hline
 & \multicolumn{3}{c}{model 4} & \multicolumn{3}{c}{model 6} \\

 & D & N & B & D & N & B \\
 \hline
 \texttt{zxipcf} & \multicolumn{6}{c}{{\bf warm absorber}} \\
 $N_{\rm H}\,[\times 10^{22}]$ & $39.2^{+40.4}_{-19.5}$ & $25.7^{+8.7}_{-8.6}$ & $36.6^{+29.6}_{-24.6}$ & $12.0^{+1.3}_{-0.7}$ & $10.5^{+3.2}_{-2.3}$ & $25.4^{+32.6}_{-2.1}$ \\[0.1cm]
 $\log_{10}(\xi)$ & $2.6^{+0.4}_{-0.3}$ & $2.4^{+0.2}_{-0.1}$ & $0.4^{+0.5}_{-0}$ & $0.5^{+0.2}_{-0.1}$ & $2.6^{+0.2}_{-0.1}$ & $0.4^{+0.2}_{-0}$ \\[0.1cm]
 $f_{\rm cov}$ & $0.28^{+0.14}_{-0.10}$ & $0.22^{+0.06}_{-0.05}$ & $0.18^{+0.05}_{-0.07}$ & $0.98^{+0.02}_{-0.01}$ & $0.3^{+0.1}_{-0.1}$ & $0.21^{+0.02}_{-0.01}$ \\[0.1cm]
 $z$ & $-0.041^{+0.004}_{-0.004}$ & $-0.005^{+0.003}_{-0.005}$ & $-0.02^{+0.02}_{-0.03}$ & $0.003^{+0}_{-0.004}$ & $-0.055^{+0.001}_{-0.003}$ & $-0.02^{+0.01}_{-0.02}$ \\[0.1cm]
 \hline
 \texttt{zxipcf} & \multicolumn{6}{c}{{\bf warm absorber}} \\
 $N_{\rm H}\,[\times 10^{22}]$ & $1.7^{+3.3}_{-0.8}$ & $3.0^{+1.2}_{-0.8}$ & $3.1^{+3.1}_{-0.9}$ & - & $3.4^{+0.8}_{-0.3}$ & $3.0^{+1.1}_{-1.4}$\\[0.1cm] 
 $\log_{10}(\xi) $ & $3.1^{+0.1}_{-0.2}$ & $3.9^{+0.1}_{-0.1}$ & $3.3^{+0.1}_{-0.1}$ & - & $3.4^{+0.1}_{-0.1}$ & $3.6^{+0.3}_{-0.2}$ \\[0.1cm]
 $f_{\rm cov}$ & $1.0^{+0}_{-0.6}$ & $1.0^{+0}_{-0.1}$ & $1.0^{+0}_{-0.5}$ & - & $1.0^{+0}_{-0.06}$ & $0.7^{+0.3}_{-0.1}$ \\[0.1cm] 
 $z$ & $-0.071^{+0.005}_{-0.006}$ & $0.003^{+0}_{-0.001}$ & $-0.004^{+0.003}_{-0.002}$ & - & $0.002^{+0.001}_{-0.002}$ & $-0.001^{+0.004}_{-0.003}$ \\[0.1cm]
 \hline
 \texttt{partcov*ztbabs} & \multicolumn{6}{c}{{\bf neutral absorber}} \\
 $N_{\rm H}\,[\times 10^{22}]$ & $15.2^{+1.3}_{-2.0}$ & $13.1^{+1.1}_{-1.2}$ & $9.8^{+0.5}_{-2.9}$ & - & $14.7^{+0.6}_{-0.5}$ & $9.7^{+0.3}_{-0.7}$ \\[0.1cm]
 $f_{\rm cov}$ & $0.92^{+0.07}_{-0.01}$ & $0.99^{+0.01}_{-0.02}$ & $0.90^{+0.04}_{-0.06}$ & - & $0.96^{+0.01}_{-0.02}$ & $0.91^{+0.02}_{-0.02}$ \\[0.1cm]
 \hline
 \texttt{partcov*ztbabs} & \multicolumn{6}{c}{{\bf neutral absorber}} \\
 $N_{\rm H}\,[\times 10^{22}]$ & $131^{+43}_{-44}$ & $46.0^{+8.7}_{-11.0}$ & $13.8^{+7.4}_{-6.0}$ & - & $73.7^{+5.4}_{-4.9}$ & $13.7^{+5.1}_{-2.0}$ \\[0.1cm]
 $f_{\rm cov}$ & $0.49^{+0.09}_{-0.12}$ & $0.22^{+0.05}_{-0.05}$ & $0.25^{+0.18}_{-0.02}$ X12 & - & $0.32^{+0.02}_{-0.01}$ & $0.25^{+0.09}_{-0.02}$ X12 \\[0.1cm]
 & & & $0.05^{+0.24}_{-0}$ X3 & & & $0.05^{+0.11}_{-0}$ X3 \\[0.1cm]
 \hline
 \texttt{reflkerr} & \multicolumn{6}{c}{{\bf disk reflection}} \\
 $\Gamma$ & $1.67^{+0.05}_{-0.05}$ & $1.73^{+0.01}_{-0.01}$ & $1.74^{+0.03}_{-0.03}$ & $1.80^{+0.08}_{-0.04}$ & $1.90^{+0.01}_{-0.01}$ & $1.78^{+0.01}_{-0.03}$ \\[0.1cm]
 $kT_{e}$ & $44^{+32}_{-8}$ & $39^{+5}_{-7}$ & $42^{+7}_{-5}$ & $373^{+94}_{-109}$ & $320^{+7}_{-14}$ & $49^{+6}_{-5}$\\[0.1cm]
 $i \; [^{\circ}]$ & - & \multicolumn{2}{c}{$20.5^{+1.7}_{-2.3}$(l)} & - & \multicolumn{2}{c}{$10.4^{+1.2}_{-0.4}$(l)} \\[0.1cm]
 $Z_{\rm Fe}$ & - & \multicolumn{2}{c}{$0.65^{+0.08}_{-0.06}$(l)} &  \multicolumn{3}{c}{$4.2^{+0.1}_{-0.1}$(l)} \\[0.1cm]
 $\log_{10}(\xi_{\rm disk}) $ & - & $2.6^{+0.1}_{-0.1}$ & $3.0^{+0.1}_{-0.6}$ & - & $2.6^{+0.4}_{-0.1}$ & $2.9^{+0.2}_{-0.2}$ \\[0.1cm]
 $r_{\rm in}\; [r_{\rm g}]$ & - & $8.1^{+1.4}_{-0.9}$ & $60^{+85}_{-31}$ & - & $7.1^{+4.7}_{-3.1}$ & $56^{+12}_{-7}$ \\[0.1cm] 
 $\mathcal{R_{\rm disk}}$ & - & $0.25^{+0.03}_{-0.02}$ & $0.14^{+0.07}_{-0.05}$ & - & $0.02^{+0.01}_{-0.01}$ & $0.03^{+0.01}_{-0.01}$ \\[0.1cm]
 $N \; [\times 10^{-2}]$ & $3.3^{+0.8}_{-0.6}$ & $6.5^{+0.5}_{-0.2}$ & $8.8^{+1.2}_{-1.0}$ & $2.5^{+0.6}_{-0.5}$ & $12^{+3}_{-4}$ & $10^{+1}_{-1}$ \\[0.1cm]
 \hline
 \texttt{rdblur*hreflect} & \multicolumn{6}{c}{{\bf distant reflection}} \\
 $i \; [^{\circ}]$ & - & - & - & \multicolumn{3}{c}{$2.8^{+1.1}_{-0.4}$(l)} \\[0.1cm]
 $r_{\rm in}\; [r_{\rm g}]$ & - & - & - & $1000^{+0}_{-261}$ & $316^{+95}_{-68}$ & $1000^{+0}_{-555}$ \\[0.1cm] 
 $\mathcal{R_{\rm dist}}$ & - & - & - & $0.70^{+0.05}_{-0.02}$ & $0.22^{+0.01}_{-0.01}$ & $0.13^{+0.01}_{-0.01}$ \\[0.1cm]

 \hline
 \texttt{zgauss} & \multicolumn{6}{c}{{\bf narrow Fe K$\alpha$ line}} \\
 $E_\alpha \;[\rm keV]$ & $6.39^{+0.01}_{-0.01}$ & $6.38^{+0.01}_{-0.01}$ & $6.41^{+0.02}_{-0.01}$ & - & - & - \\[0.1cm]
 $\sigma\;[\rm eV]$ & $19^{+9}_{-9}$ & $22^{+12}_{-14}$ & $55^{+21}_{-22}$ & - & - & - \\[0.1cm]
 $N_\alpha \; [10^{-4}$ph/(cm$^2$s)$]$ & $3.4^{+0.5}_{-0.5}$ & $2.9^{+0.1}_{-0.1}$ & $2.1^{+0.6}_{-0.4}$ & - & - & - \\[0.1cm]

 \hline
 \texttt{compps} & \multicolumn{6}{c}{{\bf soft component}} \\
 $N \; [\times 10^{-3}]$ & $<0.96$ & $2.9^{+0.1}_{-0.6}$ & $2.7^{+0.3}_{-1.7}$ & $<0.78$ & $2.9^{+0.1}_{-0.1}$ & $2.3^{+0.7}_{-1.3}$ \\[0.1cm]
 \hline
 $\chi^{2}/{\rm DoF}$ & \multicolumn{3}{c}{1790/1573} & \multicolumn{3}{c}{1852/1583} \\[0.1cm]
 $\chi^{2}/{\rm DoF}$ & 528/410 & 577/458 & 685/703 & 557/417 & 605/458 & 690/703 \\
 \hline
\end{tabularx}%
}\\
   \vskip 0.05cm
{\it Notes:} The relativistic reflection component in D is neglected due to its negligible significance. Parameters denoted with '(l)' are linked across the three spectra; the DoF for the individual spectra include these parameters, therefore, their sum is larger than the DoF for the joint fit. In model 6, $Z_{\rm Fe}$ in \reflkerr and in \hreflect are linked, also $i$ in \hreflect and in \texttt{rdblur} are linked (because the light bending as well as relativistic aberration are negligible at the fitted distances). $\xi$ is given in the unit of ${\rm erg\; cm\; s}^{-1}$. See the Notes of Table \ref{tab:1} for definition of $N$ and description of the soft component.
 \label{tab:model456}
\end{table*}

The relativistic reflection parameters appear to be reliably constrained for spectrum N. Allowing $q$ to vary for that spectrum  we find $q = 3.1^{+0.6}_{-0.3}$ and the fitted value of $r_{\rm in}$ is not affected, i.e.\ $r_{\rm in}=8.0^{+1.4}_{-1.3}$. In turn, the reflection parameters are poorly constrained for spectrum B, for which any value of $2 \le q \le 10$ gives the same value of $\chi^2$. However, for B we also note a correlation between $q$ and $r_{\rm in}$, e.g.\ we get $r_{\rm in}=31^{+128}_{-29}$ for (fixed) $q=2$ and $r_{\rm in}=77^{+78}_{-31}$ for $q=5$, which disfavors a significant contribution from $r \la 50$.

In model 6, the (weakly blurred) Compton reflection continuum, produced together with the narrow Fe K line, reduces a potential contribution of the inner disk reflection to $\mathcal{R}_{\rm disk} \la 0.05$ and the significance of the inner disk reflection is then only marginal, namely setting $\mathcal{R}_{\rm disk}=0$ gives $\Delta \chi^2 = 6$ for N and $\Delta \chi^2 = 9$ for B. For spectrum D we again found no significant indication for the inner disk reflection ($\Delta \chi^2 = -3$ for 3 degrees of freedom), therefore, we neglected it. 

\begin{figure*}[htp]
  \centering
    \includegraphics[scale=0.7,clip]{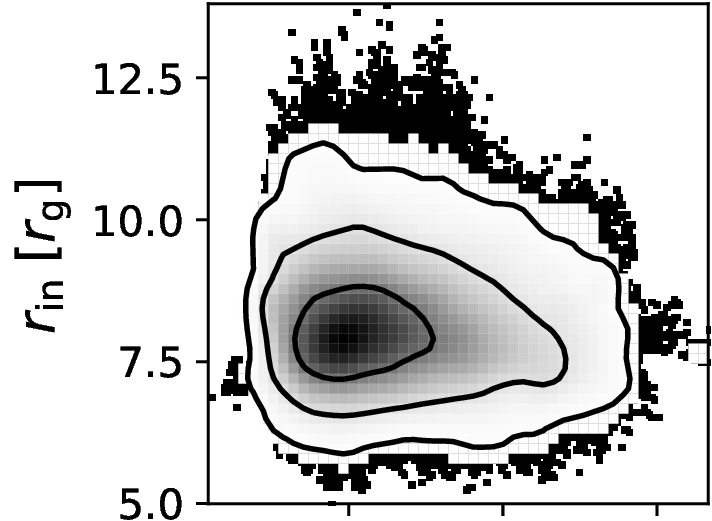}
    \includegraphics[scale=0.7,clip]{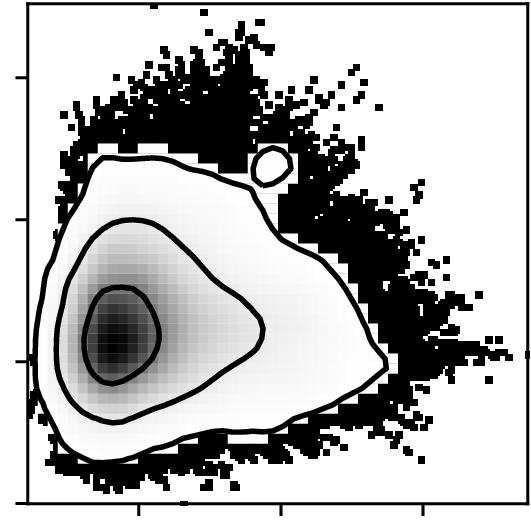}
    \includegraphics[scale=0.7,clip]{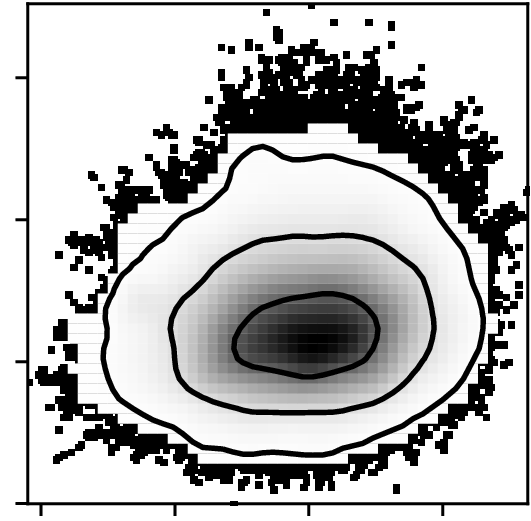}
    \includegraphics[scale=0.7,clip]{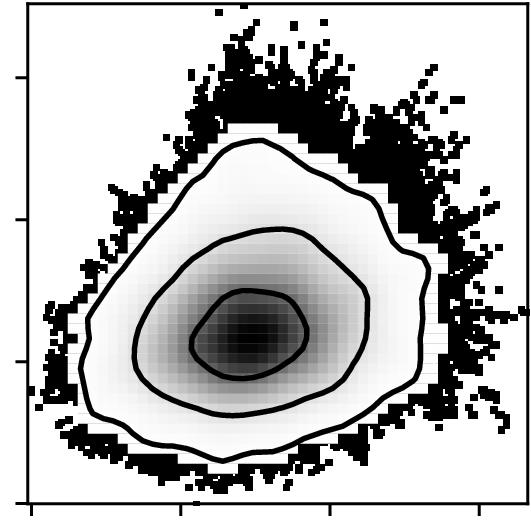}\\
    \includegraphics[scale=0.7,clip]{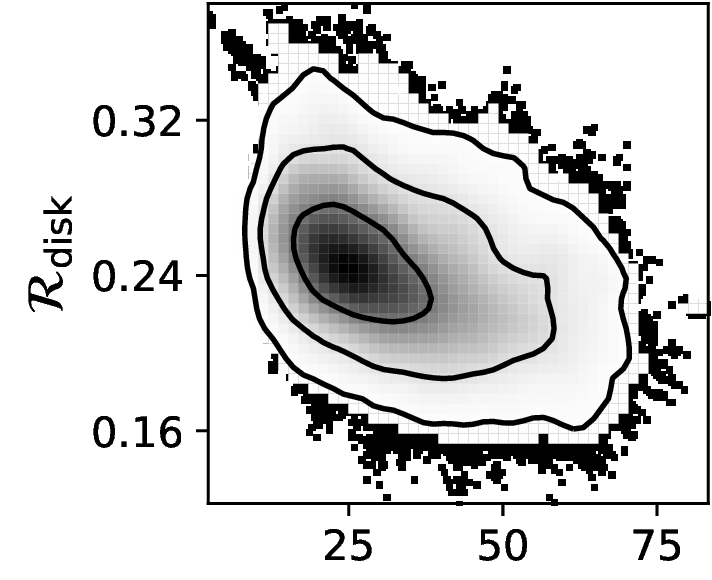}
    \includegraphics[scale=0.7,clip]{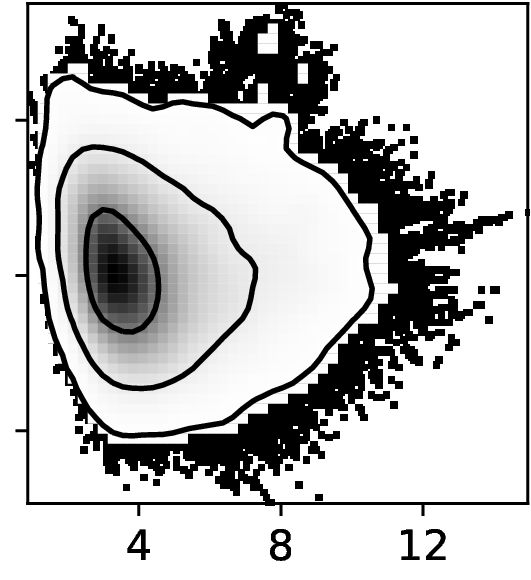}
    \includegraphics[scale=0.7,clip]{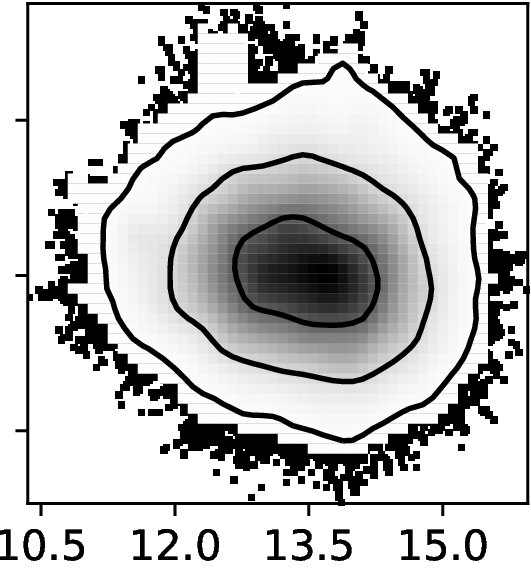}
    \includegraphics[scale=0.7,clip]{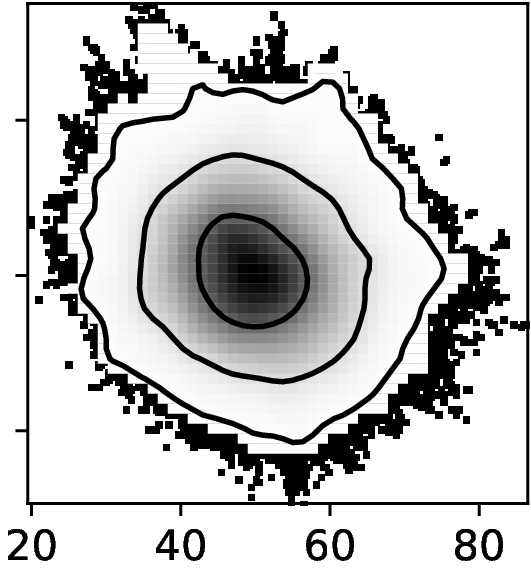}\\
    \hspace{1.5cm}$N_{\rm H}\;[\times 10^{22}\;{\rm cm}^{-2}]$
    \vspace{.5cm}
\caption{Probability distributions showing correlations between $r_{\rm in}$ and $N_{\rm H}$ (top panels) and $\mathcal{R_{\rm disk}}$ and $N_{\rm H}$ (bottom panels), in the model 4 for spectrum N, obtained in the MCMC analysis using \texttt{xspec\_emcee} implemented by Jeremy Sanders. The columns from left to right correspond to the absorption components from top to bottom in Table \ref{tab:model456}. The contours correspond to the significance of $\sigma = 1, 2, 3$.}
  \label{fig:corr}
\end{figure*}

We did not find any significant degeneracies between the parameters of absorption and reflection components. Figure \ref{fig:corr} shows example correlation plots of $r_{\rm in}$ and $\mathcal{R_{\rm disk}}$ against $N_{\rm H}$ of the four absorption components in the model 4 for spectrum N, obtained using the Markov Chain Monte Carlo (MCMC) method. The correlation plots for spectra B and D are qualitatively similar to these in Figure \ref{fig:corr}.

\section{Discussion}
\label{sect:discuss}

Table \ref{tab:summary} compares the likelihoods of the lamp-post and the truncated disk models. Model 5 is almost the same as model 6, except for the lack of relativistic blurring of distant reflection and the related increase of $\chi^2$, therefore, it is not included. The likelihoods are found using the Akaike information criterion \citep{1973Akaike}, with a correction for the finite size of sample, strictly following the procedure described in \citet{2019MNRAS.485.3845D}. We see that the model 4 is strongly favored. The alternative models are ruled out for spectrum N as well as for the joint fits, and they have low likelihoods for spectra B and D.

Application of our \reflkerrlp model to the \suzaku/\nustar data for \source rules out the lamp-post geometry proposed for these data by \citet{keck15} and \citet{beuchert17}, since a very poor fit is obtained with the properly calculated Comptonization spectrum. This demonstrates the importance of a proper modeling of gravitational redshift acting on the direct radiation from a source close to the horizon, i.e., at the location needed to explain the apparently very broadened reflection components. This redshift cannot be compensated by an increase of electron temperature, even in the data extending only up to $\simeq 80$ keV. \citet{2016ApJ...821L...1N} discussed further problems with this geometry related with trapping of a majority of X-ray photons under the horizon as well as extreme values of the compactness parameter, $\ell$, implying a runaway e$^\pm$ pair production for $\ell \gg 10^3$. For the LP$_{1}$ component found in our best fitted lamp-post model (model 3), the trapping reduces the \xray luminosity by a factor of 100 (spectrum D, $h=1.17$), 25 (spectrum N, $h=1.33$) and 10 (spectrum B, $h=1.6$), and the corresponding $\ell \simeq 2 \times 10^6$, $3 \times 10^5$ and $10^4$, respectively. Yet another problem for the lamp-post model follows from the the \xray/UV/optical delay pattern measured in \source \citep{2017ApJ...840...41E}, which cannot be explained by reprocessing in the disk strongly illuminated by a compact lamp-post.

We found that models with a truncated disk give a much better spectral description for \source. Here the inferred properties of the inner flow depend on the nature of the distant reprocessor producing the narrow Fe K line. This distant matter appears to be not very distant but rather located at a distance of several hundred $R_{\rm g}$, as indicated by our modeling of the narrow line. We agree in this point with other recent studies of NGC 4151. Our measured Fe K$\alpha$ line widths of $\sim 20$ eV are consistent with those obtained from the higher-resolution {\it Chandra} data \citep{2018ApJ...865...97M}, where also the line asymmetry was found indicating the presence of a redshifted line component and parameters of the line-emitting region similar to ours ($r_{\rm in}$ of several hundred and very low $i$) were estimated. \citet{2019ApJ...884...26Z} report variability of the narrow Fe K$\alpha$ line and a measurement of a delay of $\sim 3$ days in its response to the \xray continuum, which corresponds to the distance of $\sim 10^3 R_{\rm g}$ for $M=5 \times 10^7 M_\odot$. This is larger by a factor of $\sim$3 than $r_{\rm in}$ of the line production site which we found for data set N. However, \citet{2019ApJ...884...26Z} estimated the 3-day delay basing primarily on the \xmm data, whereas the \suzaku observations alone (the same as in our data sets D and N) indicate a much shorter delay of $\la$1 day (see their figure 10d). 

Our overall best fit is obtained in the model assuming that the source of the narrow Fe K$\alpha$ line is Compton thin, so that the Fe line is not accompanied by a strong Compton-reflected continuum. In this scenario, it is likely that the narrow Fe K$\alpha$ line is produced in the material responsible for the neutral absorption, most likely by the clouds in the broad line region (BLR). We can then compare the fitted $N_{\rm H}$ with the column densities needed to produce the line with the measured EW, although we note that our absorption fits give the line-of-sight $N_{\rm H}$, whereas the line production efficiency depends on the angle averaged value. Furthermore, the theoretical EW depends on many factors, in particular the geometry, orientation and covering factor of the line-emitting material \citep{2010MNRAS.401..411Y}, which are very uncertain. In the context of the model of \citet{2009ApJ...692..608I}, the ${\rm EW} \simeq 220$ eV (D), 100 eV (N) and 60 eV (B) of the narrow Fe K$\alpha$ line with respect to the main Comptonization component in our Compton-thin model (model 4), may be produced at $N_{\rm H} \simeq (1$--$3) \times 10^{23}$ cm$^{-2}$. These values are indeed approximately consistent with $N_{\rm H}$ of the neutral absorber in the model 4, where also the decrease of the fitted $N_{\rm H}$ between D and B is in agreement with the measured decrease of EW. The weakly covering neutral absorber, fitted with $f_{\rm cov} \la 0.5$, indicates even higher $N_{\rm H}$, but this component likely describes small clouds passing in the light-of-sight (as partial covering in time is more likely than in space here) so its $N_{\rm H}$ may not be representative for angle-averaged column densities. 

Models assuming that the narrow Fe K line is produced in a Compton thick matter give a worse spectral description of the \xray data. Interestingly, the \hreflect components are fitted with similar normalizations in all spectral sets in these models (see Figure \ref{fig:q3}b for model 6, the spectral decomposition for model 5 is similar). Such a weakly variable reflection component could be expected for reflection produced at a parsec scale (e.g.\ in a molecular torus postulated in the AGN unified model) and then averaged over different emission states of the central source. However, the reflection appears to take place mostly on a much closer spatial scale, as discussed above. On the other hand, the fitted values of $\mathcal{R_{\rm dist}}$ challenge the model with reflection from the intermediate region of a warped disk, proposed by \citet{2018ApJ...865...97M}. Here, a significant change of the solid angle subtended by the disk at the \xray source, with $\Omega \simeq 1.4 \pi$ in D, $0.4 \pi$ in N and $0.2 \pi$ in B, seems very unlikely (and we did not find any trends in the {\it Swift}/BAT light curves during several-day periods preceding observations D and N, which could explain the reflection steadiness by flux variability).

Describing the narrow Fe K$\alpha$ line with the disk emission model, \texttt{rdblur} or \texttt{diskline}, we obtained inclinations formally indicating a strictly face-on orientation of the disk. However, if the line is emitted from the BLR, this fitted line profile more likely indicates a significant contribution from a non-disk component, whose chaotic motion smears the disk-like emission line. Such a complex configuration of the BLR is indicated by its formation models \citep[e.g.][]{2011A&A...525L...8C} and by the observed profiles of the optical emission lines \citep[e.g.][]{2009MNRAS.400..924B}.

The inclination angle of the fitted inner reflection model gives in turn the actual inclination of the accretion disk. We found small $i$, in agreement with all previous applications of relativistic reflection models to NGC 4151 \citep[e.g.][]{2002ApJ...573..505Z,keck15,beuchert17}, where even lower values than $i \simeq 20 \degr$ of our model 4 were typically found. This low $i$ is seemingly in conflict with a strong obscuration of the X-ray spectrum in NGC 4151, which according to the original unification scheme of AGNs \citep[e.g.][]{1993ARA&A..31..473A} should correspond to rather large inclination with the line-of-sight crossing the molecular torus. In our current view of AGN unification, however, it has become clear that additional components should be incorporated, in particular obscuration by gas in the BLR \citep[e.g.][]{2015ARA&A..53..365N}, which makes obscuration not being due to orientation effects only (as proposed in the simplest unification model). This is likely the case for NGC 4151, where obscuration is observed to be highly variable and then it occurs on a spatial scale close to the central region.

The electron temperature of the X-ray source is $\simeq 40$ keV in model 4 fitted to all three spectral sets. In model 6, $kT_{\rm e} \simeq 40$ keV in B and $\simeq (300$--400) keV in D and N. This difference between models 4 and 6 results from a strong contribution of the neutral reflection needed to fit the soft X-ray part of spectrum D in the latter model, which then requires a large $T_{\rm e}$ to reproduce the hard X-ray part by the total model spectrum. If no reflection component is present (as in model 4 for D) the same data can be reproduced by the thermal Comptonization spectrum with much lower $T_{\rm e}$. The difference in $T_{\rm e}$ fitted to spectrum N in models 4 and 6 results from the difference of $\Gamma$ in these models.
We note that a high value of $kT_{\rm e} \simeq 320$ keV is precisely determined in the model 6 fit to spectrum N despite using the data extending only up to 80 keV (i.e.\ not covering the high energy cut-off produced at this temperature) due to the dependence of the Compton reflection spectrum on $T_{\rm e}$ at these energies.

Also $N_{\rm H}$ of the absorbing material estimated for spectrum D shows a strong dependence on the assumed spectral model. The soft X-ray part of this spectrum can be described by either a strong ($\mathcal{R} \simeq 0.7$), neutral reflection component (as in model 6), or a neutral material with $\tau \sim 1$, absorbing $\sim 50$\% of the X-ray emission (as in model 4). The latter solution has a higher statistical likelihood. It also fits in a scenario in which the column density of the neutral absorber increases with decreasing hard X-ray flux (with the total $f_{\rm cov} N_{\rm H}/10^{22} \, {\rm cm}^2 \simeq 10$ in B, 20 in N and 80 in D). No such a systematic trend is found in the former solution (with a strong neutral reflection).

Altogether, our results indicate that the disk in \source is truncated, which favors the geometry with X-rays produced by an inner hot flow. In our best spectral model the truncation occurs relatively close to the black hole (within a $100R_{\rm g}$) when the source is bright in hard X-rays, and reflection from the inner disk is weak, $\mathcal{R_{\rm disk}} \sim 0.2$, which implies that the vertical scale height of the flow is small. Our best model is consistent with the spectral-timing model proposed for \source by \citet{2020MNRAS.491.5126M}, where the main reprocessing of the \xray radiation occurs in optically thin clouds on size scales of the BLR. The presence of the inner disk estimated in our analysis does not conflict with the dominating reprocessing in the BLR, because only a small fraction of X-rays is reprocessed in the disk.

\section{Conclusions}

We have found that the lamp-post model proposed in previous works for NGC 4151 gives a poor spectral description when GR effects are properly included in the computation of the thermal Comptonization spectrum. Much better spectral solutions are obtained by allowing the optically-thick disk to be truncated. Our overall best solution indicates that the narrow Fe K$\alpha$ line is produced by Compton-thin matter in the BLR, which is similar to the well-established case of another bright Seyfert 1 galaxy, NGC 7213 \citep{2008MNRAS.389L..52B}. In this solution we find also a weak but significant reflection from a disk truncated at $\sim$(10--$60) R_{\rm g}$ when the source is in bright emission states. We did not find evidence either for or against the presence of the disk when the source is in the dim state. We also did not find any significant evidence for reflection from the inner disk in the alternative model, with the narrow Fe K$\alpha$ line produced by reflection from the disk, at $\ga 300 R_{\rm g}$ during the \nustar observation and at larger distances in the remaining data sets. This alternative model is disfavored, however, by its statistical inferiority as well as physical consistency problems, most notably large changes of the angle subtended by the reflector (presumably the intermediate disk) as seen by the X-ray source, required to explain the reflection amplitude. 

\section*{ACKNOWLEDGEMENTS}
We thank the anonymous referee for helpful comments, Tobias Beuchert for a clarification regarding his model and Piotr Lubi\'nski for providing the spectral data. This research has been supported in part by the Polish National Science Centre grants 2015/18/A/ST9/00746, 2016/21/B/ST9/02388 and 2019/35/B/ST9/03944. A.A.Z.\ and A.N.\ are members of International Teams at the International Space Science Institute (ISSI), Bern, Switzerland, and thank ISSI for the support during the meetings.


\bibliography{ngc4151_apj}{}
\bibliographystyle{aasjournal}

\label{lastpage}
\end{document}